\documentclass{article}

\usepackage{INTERSPEECH2022}
\usepackage{amsmath}
\usepackage{hyperref}
\usepackage{enumitem}
\usepackage{graphicx}
\usepackage{multirow}
\usepackage{multicol}
\usepackage{xcolor}

\usepackage{soul}
\usepackage{float}
\usepackage{tabularx}
\usepackage{csvsimple}
\usepackage{mathtools}
\usepackage{amssymb}
\usepackage{commath}
\usepackage{subcaption}
\usepackage{cite}
\usepackage{hyperref}
\usepackage{relsize}
\usepackage{romannum}
\usepackage{bm}
\usepackage{stmaryrd}

\def\vec#1{\ensuremath{\bm{{#1}}}}

\title{CopyCat2: A Single Model for Multi-Speaker TTS and Many-to-Many Fine-Grained Prosody Transfer}
\begin{document}
	\name{
		Sri Karlapati, 
		Penny Karanasou, 
		Mateusz Lajszczak, 
		Ammar Abbas, 
		Alexis Moinet, 
		Peter Makarov$^*$, 
		Ray Li$^*$, 
		Arent van Korlaar, 
		Simon Slangen$^*$\thanks{$^*$Work done while at Amazon.},
		Thomas Drugman
	}

	\address{
		Alexa AI, Amazon, Cambridge, United Kingdom
	}
	
	\email{
	    srikarla@amazon.com
	}

	\maketitle

	\begin{abstract}
	    In this paper, we present CopyCat2 (CC2), a novel model capable of: a) synthesizing speech with different speaker identities, b) generating speech with expressive and contextually appropriate prosody, and c) transferring prosody at fine-grained level between any pair of seen speakers. We do this by activating distinct parts of the network for different tasks. We train our model using a novel approach to two-stage training. In Stage~I, the model learns speaker-independent word-level prosody representations from speech which it uses for many-to-many fine-grained prosody transfer. In Stage~II, we learn to predict these prosody representations using the contextual information available in text, thereby, enabling multi-speaker TTS with contextually appropriate prosody. We compare CC2 to two strong baselines, one in TTS with contextually appropriate prosody, and one in fine-grained prosody transfer. CC2 reduces the gap in naturalness between our baseline and copy-synthesised speech by $22.79\%$. In fine-grained prosody transfer evaluations, it obtains a relative improvement of $33.15\%$ in target speaker similarity.
	\end{abstract}
	\section{Introduction}
	    \label{sec:introduction}
		Neural Text-to-Speech (NTTS) techniques have significantly improved the naturalness of speech produced by TTS systems\cite{tacotron2, UniversalWavenet, li2018close, lorenzo2018robust, zhang2019learning, skerry2018towards, CAMP, Kathaka, tyagi2020dynamic}. In order to improve the prosody of TTS, there has been considerable work in learning prosody representations from ground-truth speech\cite{zhang2019learning, skerry2018towards, CAMP, Kathaka, tyagi2020dynamic}. These representations are usually learnt using encoders at various time scales and are used in conjunction with phonemes to generate speech using either an auto-regressive or parallel decoder. As per the subtractive definition of prosody\cite{skerry2018towards}, since phonetic information is readily available to the decoder, it will rely on the additional encoders to learn prosody features. During inference, these prosody representations are unavailable. Therefore, in order to generate speech with contextually relevant prosody, some approaches learn to predict these prosody representations using textual features and use them in place of the prosody representations\cite{CAMP, Kathaka,tyagi2020dynamic,klapsas2021word, guo2022unsupervised, ren2021portaspeech, hayashi2019pre}. These methods, however, mainly focus on single-speaker data, and cannot trivially be made into multi-speaker models as it would require changes to how these representations are learnt \cite{karanasou21_interspeech}.
		
		Another extension to NTTS is prosody transfer (PT) \cite{wang2018style}, where the prosody from a source audio is used as a reference when synthesising speech in a target speaker's voice. Fine-Grained Prosody Transfer (FPT) techniques \cite{zhang2018learning, lee2018robust, klimkov2019finegrained, karlapati2020copycat} mainly focus on capturing temporal prosody features like rhythm, emphasis, loudness, and melody. FPT methods aim to learn speaker-independent prosody representations from ground-truth speech using an encoder, and then in conjunction with phonemes and speaker representations, learn to predict the same speech using a decoder. During inference, a different target speaker representation is provided to the decoder, which then generates speech with the prosody of the reference and the voice identity of the target speaker. FPT techniques can suffer from speaker leakage, where the speech generated has the source speaker's identity and not the target speaker's identity \cite{karlapati2020copycat}. Most work in FPT has been on reducing speaker leakage, but little research has been done to see if the speaker-independent prosody representations learnt for FPT could be used for downstream tasks like TTS with contextually appropriate prosody.

		In this paper, we present CopyCat2 (CC2) with $4$ main contributions. First, we show through subjective evaluations that learning speaker-independent prosody representations at the word-level results in reduced speaker leakage in FPT when compared to methods like CopyCat1 (CC1) \cite{karlapati2020copycat}, which learn representations that are not aligned to linguistic units. Second, we show through our experiments that we can use the speaker-independent prosody representations learnt for FPT in the downstream task of TTS with contextually appropriate prosody. Third, we show through our novel methodology that the same model can be used for both multi-speaker TTS and many-to-many FPT. Finally, using subjective evaluations, we show that learning word-level prosody representations from multi-speaker data and predicting them using text improves the naturalness of generated speech, when compared to single-speaker methods like Kathaka \cite{Kathaka} that learn prosody representations at the sentence-level and predict them using text. 
        \vspace{-0.2cm}
	\section{CopyCat2}
		\label{sec:method}
\begin{figure*}
			\centering
            \includegraphics[width=\linewidth]{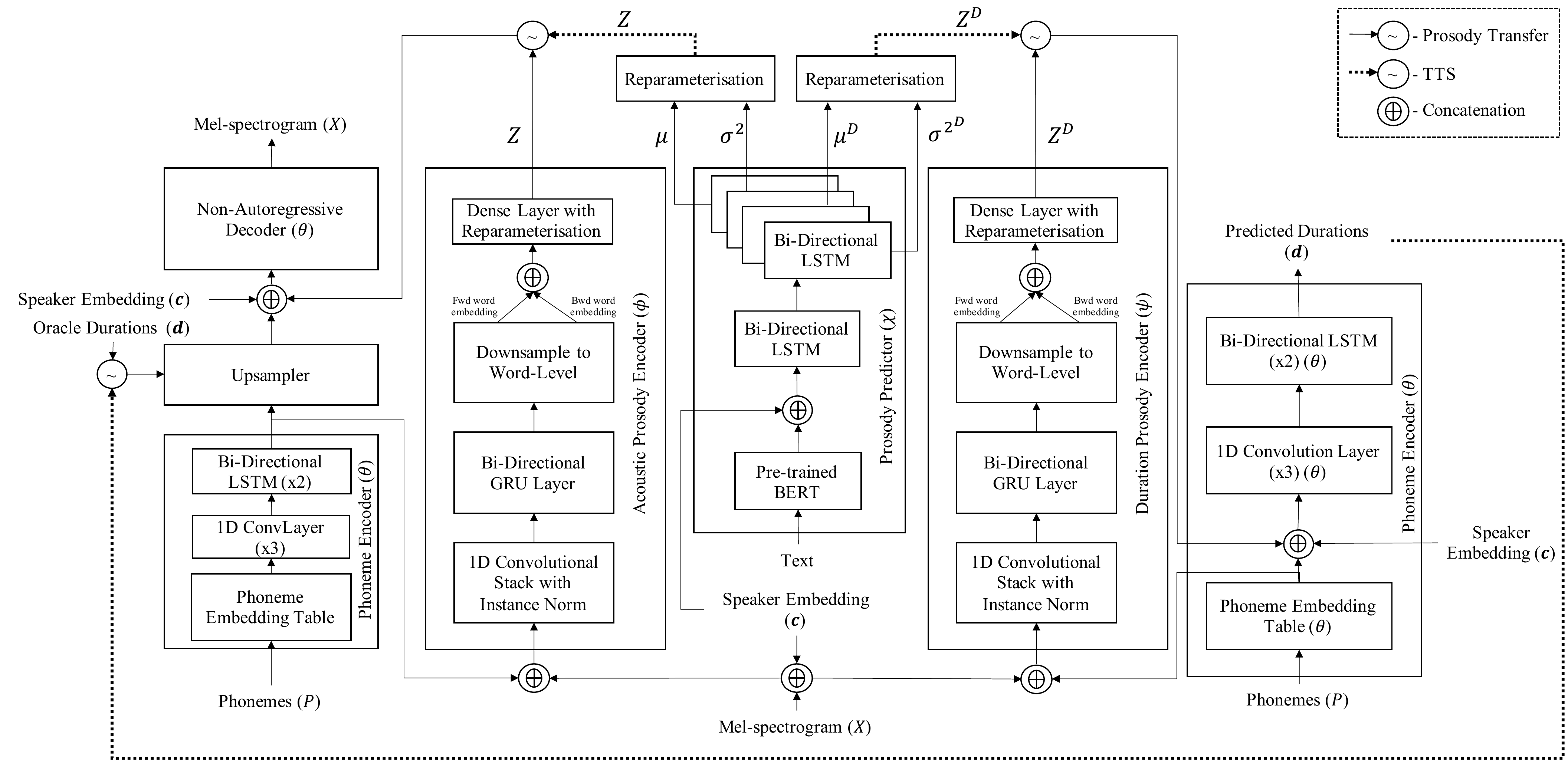}
            \vspace{-0.5cm}
			\caption{This figure consists of the acoustic model on the left, the prosody predictor in the middle, and the duration model on the right. The architecture of the acoustic and duration prosody reference encoders ($q_\phi(\cdot)$ \& $q_\psi(\cdot)$) are shown here to predict $Z$ and $Z^D$. The prosody predictor ($s_\chi(\cdot)$) is shown in the middle. Follow the hard lines to $\sim$ for FPT and the dotted lines to $\sim$ for TTS.}
			\label{fig:copycat_model}
			\vspace{-0.5cm}
		\end{figure*}

		The CC2 model consists of $3$ major components (Fig.\ref{fig:copycat_model}): 1) the acoustic model, 2) the duration model, and 3) the prosody predictor. These components are trained over two stages. In Stage~I, we learn word-level speaker-independent prosody representations from multi-speaker speech by training the acoustic and duration models using word-level conditional variational reference encoders. In Stage~II, we learn to predict these representations using the contextual information available in text using the prosody predictor. 
		\subsection{Stage~I: Learning Word-Level Speaker-Independent Prosody Representations}
		    \label{ssec:stage-I}
		    \vspace{-0.1cm}
		    \subsubsection{Acoustic Model}
		        \vspace{-0.1cm}
		        \label{sssec:acoustic_model}
		        The acoustic model of CC2 is similar to the architecture of CC1\cite{karlapati2020copycat}. It consists of a phoneme encoder, a non-autoregressive (NAR) decoder, and a conditional variational reference encoder. The phoneme encoder is similar to the character encoder in \cite{tacotron2}. It takes phonemes $\vec{y}$ as input and provides phoneme encodings $Y \in \mathbb{R}^{P \times J}$ as output, where $P$ is the number of phonemes and $J$ is the size of each phoneme encoding. These encodings are then upsampled through replication according to the per-phoneme durations, $\vec{d}$, provided as input to the upsampler. These upsampled encodings are then passed to a NAR decoder along with speaker embeddings, $\vec{c} \in \mathbb{R}^{E}$, where $E$ is the size of a speaker embedding \cite{karlapati2020copycat}. As shown in Fig.\ref{fig:copycat_model} to the left, we represent these blocks by $\theta$. 
		        
		        We introduce a sequence of latent vectors, $Z \in \mathbb{R}^{W \times  H}$, to learn word-level speaker-independent prosody representations from speech, where $W$ is the number of words in the input and $H$ is the size of each representation. Therefore, we obtain the acoustic model, $p_\theta(X \mid \vec{y}, \vec{c}, Z)$, where $X \in \mathbb{R}^{T \times 80}$ is the mel-spectrogram with $T$ frames. Since the decoder is already receiving phonetic information and speaker identity, we expect, $Z$ to encode speaker-independent prosody information. The prosody representations, $Z$, are learnt at the word-level, unlike in CC1 where they are learnt with a fixed downsampling rate, $\beta$. In order to learn the distribution of $Z$, we need to model the true posterior distribution $p_\theta(Z \mid X, \vec{y}, \vec{c})$ which is intractable as our NAR decoder is not invertible \cite{kim2020glow, miao2020flow, ren2021portaspeech, vits-kim21f,jeong21_interspeech, popov2021grad}. Therefore, we approximate the true posterior distribution using a conditional variational reference encoder, $q_\phi(Z \mid X, \vec{y}, \vec{c})$, parameterised by $\phi$. We assume, for simplicity, that the word-level prosody representations are conditionally independent of one another, $q_\phi(Z \mid X, \vec{y}, \vec{c}) = \prod_{i=0}^{W-1}q_\phi(\vec{z}_i\mid X, \vec{y}, \vec{c})$, and that the word-level speaker-independent prosody representations follow a Gaussian distribution with diagonal covariance, $q_\phi(\vec{z}_i\mid X, \vec{y}, \vec{c}) = \mathcal{N}(\vec{\mu}_i, \vec{\sigma}^2_i)$. Upon the addition of the reference encoder, we obtain the following optimisation function, which is the evidence lower bound (ELBO) to be maximised for training a Conditional~VAE \cite{sohn2015learning}:
		        \vspace{-0.2cm}
                \begin{equation}
                    \label{eq:elbo_1}
                    \begin{split}
                        L_{\phi, \theta} & = \mathop{\mathbb{E}}_{Z \sim q_\phi}[log\ p_\theta (X \mid \vec{y}, \vec{c}, Z)] \\ & - \sum_{i=0}^{W-1} KL(q_\phi(\vec{z}_i \mid X, \vec{y}, \vec{c}) \mid \mid p_\theta(\vec{z}_i \mid \vec{y}, \vec{c})).
                    \end{split}
                \end{equation}
                Training the model with this ELBO was unstable as the prior, $p_\theta(\vec{z}_i \mid \vec{y}, \vec{c})$, and posterior, $q_\phi(\vec{z}_i \mid X, \vec{y}, \vec{c})$, were trained simultaneously. To stabilise training, we replaced the prior with a simple non-parameterised prior, $p(\vec{z}_i)=\mathcal{N}(\vec{0}, \mathbb{I})\  \forall \ \vec{z}_i \in Z $, and trained to maximise the following ELBO:
                \vspace{-0.2cm}
                \begin{equation}
                    \label{eq:elbo_2}
                    \begin{split}
                        L_{\phi, \theta} & = \mathop{\mathbb{E}}_{Z \sim q_\phi}[log\ p_\theta (X \mid \vec{y}, \vec{c}, Z)] \\ & - \alpha \sum_{i=0}^{W-1} KL(q_\phi(\vec{z}_i \mid X, \vec{y}, \vec{c}) \mid \mid p(\vec{z}_i)),
                    \end{split}
                    \vspace{-0.2cm}
                \end{equation}
                where $\alpha \in [0, 1]$ is used as the annealing factor \cite{sonderby2016train} and is scaled linearly with the number of training steps. 
                
                The model learns to predict speaker-independent word-level prosody representations from multi-speaker speech, and to use this in conjunction with phonemes and speaker identity to generate mel-spectrograms. We can now perform FPT between different speakers where both the source and generated samples will have the same duration. However, since the prior cannot be sampled  using text, the model is incapable of TTS with contextually relevant prosody. In order to bypass this limitation to TTS, in Sec.\ref{ssec:stage-II}, we introduce methods that learn to predict the prosody distribution using the contextual information available in the text.
                \vspace{-.3cm}
            \subsubsection{Duration Model}
                \vspace{-.1cm}
                \label{sssec:duration_model}
                The duration model is trained to predict the number of mel-spectrogram frames each phoneme spans over, $\vec{d}=[d_0, d_1, \dots, d_{P-1}]$ \cite{yu2020durian, Kathaka, CAMP}. We obtain the target durations through forced alignment\cite{Povey_ASRU2011}, and use the same durations as oracle durations while training the acoustic model. The duration model consists of a phoneme encoder, similar to the one used in the acoustic model, as shown in Fig.\ref{fig:copycat_model}. Since the acoustic model is trained with force aligned durations, we hypothesise that the prosody representations from the acoustic model, $Z$, will capture limited duration-related prosody information. Therefore, for the remainder of the paper, we refer to $Z$ as acoustic prosody representations. To represent word-level speaker-independent duration prosody representations, we use a sequence of latent vectors, $Z^D \in \mathbb{R}^{W \times H^D}$, where $H^D$ is the size of each representation. Thereby, we obtain the duration model $r_\theta (\vec{d} \mid \vec{y}, \vec{c}, Z^D)$. To learn $Z^D$, we add a conditional variational reference encoder to the duration model, $q_\psi(Z^D \mid X, \vec{y}, \vec{c})$. We assume that mel-spectrograms capture word-level duration information, and therefore can be used in place of the target durations $\vec{d}$ for learning the posterior distribution which would otherwise be $q_\psi(Z^D \mid \vec{d}, \vec{y}, \vec{c})$. Thereby, we obtain the following ELBO to maximise:
                \vspace{-.1cm}
                \begin{equation}
                    \label{eq:elbo_dur}
                    \begin{split}
                        L_{\psi, \theta} & = \mathop{\mathbb{E}}_{Z^D \sim q_\psi}[log\ r_\theta (\vec{d} \mid \vec{y}, \vec{c}, Z^D)] \\ & - \alpha \sum_{i=0}^{W-1} KL(q_\psi(\vec{z}^D_i \mid X, \vec{y}, \vec{c}) \mid \mid p(\vec{z}^D_i)).
                    \end{split}
                \end{equation}
                We now have word-level speaker-independent duration prosody representations. These representations enable us to transfer the rhythm of speech from a source audio to a target speaker by altering the target speaker embedding. This enables FPT with the predicted durations matching the duration distribution of the target speaker and the rhythm of the source audio.
        \subsection{Stage~II: Predicting Word-Level Speaker-Independent Prosody Representations from Context}
            \label{ssec:stage-II}
            \vspace{-.05cm}
            In Stage~I, we trained acoustic and duration models which rely on acoustic prosody representations ($Z$) and duration prosody representations ($Z^D$) respectively to determine the prosody with which speech is synthesised. Since we are able to obtain these representations from a source mel-spectrogram, the model is capable of FPT. However, for the model to be capable of TTS with contextually appropriate prosody, we need to predict $Z$ and $Z^D$ using textual information. In order to do so, we introduce the prosody predictor \cite{Kathaka, CAMP}.
            \vspace{-.2cm}
            \subsubsection{Prosody Predictor}
                \label{sssec:prosody_predictor}
                \vspace{-.1cm}
                We exploit the contextual information available in a sentence to predict its prosody, since it has been cogently hypothesised that prosody is driven by it \cite{wagner2010experimental}. The acoustic and duration prosody encoders predict the parameters of a Gaussian distribution, $\mathcal{N}(\vec{\mu}_i, \vec{\sigma}_i^2)$ and $\mathcal{N}(\vec{\mu}_i^D, {\vec{\sigma}_i^D}^2)$, for each word in a sentence, indexed by $i$. Therefore, given a function capable of predicting the same parameters from text, we can treat this as a distribution matching problem. We call such functions as \textit{prosody predictors}, $s_\chi(Z, Z^D \mid tx, \vec{c})$, where $\chi$ is the set of parameters and $tx$ is the input text. The prosody predictor is trained to minimise the following loss, $L_\chi$:
                \vspace{-.1cm}
                \begin{equation}
                    \label{eq:prosody_predictor}
                    \begin{split}
                        L_{\chi} = & \sum_{i=0}^{W-1} \Big( KL\big(s_\chi(\vec{z}_i \mid tx, \vec{c}) \mid \mid q_\phi(\vec{z}_i \mid X, \vec{y}, \vec{c})\big) \\ 
                        & + KL\big(s_\chi(\vec{z}_i^D \mid tx, \vec{c}) \mid \mid q_\psi(\vec{z}_i^D \mid X, \vec{y}, \vec{c})\big)\Big).
                    \end{split}
                \end{equation}
                BERT\cite{devlin2019bert} captures contextual information in a given sentence in its embeddings\cite{rogers-etal-2020-primer}. We use a BERT model pre-trained on Wikipedia data, to obtain contextual word-piece embeddings which were then passed to 4 different LSTMs, each predicting one of the word-level prosody distribution parameters, $\vec{\mu}_i,\ \vec{\sigma}_i^2,\ \vec{\mu}_i^D$, and ${\vec{\sigma}_i^D}^2$. We concatenate speaker embeddings to the BERT embeddings before passing them to the LSTM heads, as we noted that the trajectory of consecutive prosody representations was speaker-dependent, though the representations themselves were speaker-independent. Since, we can now predict word-level prosody representations from text, we have assuaged the limitations introduced by assuming an isotropic Gaussian prior for the reference encoders in Stage~I. We use these text-predicted prosody representations in the acoustic and duration models for TTS with contextually appropriate prosody.
                \vspace{-.2cm}
		\subsection{Training \& Inference}
		    \vspace{-.1cm}
			\label{ssec:training_setup}
			\subsubsection{Training}
			    \vspace{-.1cm}
			    In both training stages, we use multi-speaker data. In Stage~I, we train the acoustic and duration models to predict mel-spectrograms and durations respectively, while also learning acoustic and duration prosody representations. Before training the prosody predictor, we run the prosody reference encoders over the training set of all the speakers in the dataset to obtain the speaker-independent word-level acoustic and duration prosody distribution parameters. These parameters are used as targets to train the prosody predictor in Stage~II.
			    \vspace{-.15cm}
			\subsubsection{Inference}
			    \vspace{-.1cm}
			    There are two modes of inference that are possible with CC2: 1) FPT and 2) TTS. At the end of Stage~I of training, CC2 is capable of FPT. We can take a mel-spectrogram from a source speaker for reference prosody. It is given as input to the duration model while providing it with a target speaker embedding. Thereby, we obtain durations in the target speaker's duration distribution but with the rhythm of the reference audio. The predicted durations are used as input to the acoustic model, along with the reference mel-spectrogram, and the target speaker embedding, to generate speech with the target speaker's identity and the acoustic prosody and rhythm of the reference audio.
			    
			    At the end of Stage~II, CC2 is capable of TTS. We pass the text and the target speaker embedding to the prosody predictor, to obtain word-level acoustic and duration prosody representations. The duration prosody representations, phonemes and the target speaker embedding, are used to predict durations that are contextually appropriate. These durations, the acoustic prosody representations, phonemes, and the target speaker embedding, are given to the acoustic model, to generate speech with the target speaker's identity and contextually appropriate prosody.
		    	\vspace{-.15cm}
	\section{Experiments}
		\label{sec:experiments}
		\begin{figure*}
			\centering
			\begin{subfigure}{0.33\linewidth}
	        	\includegraphics[width=\linewidth]{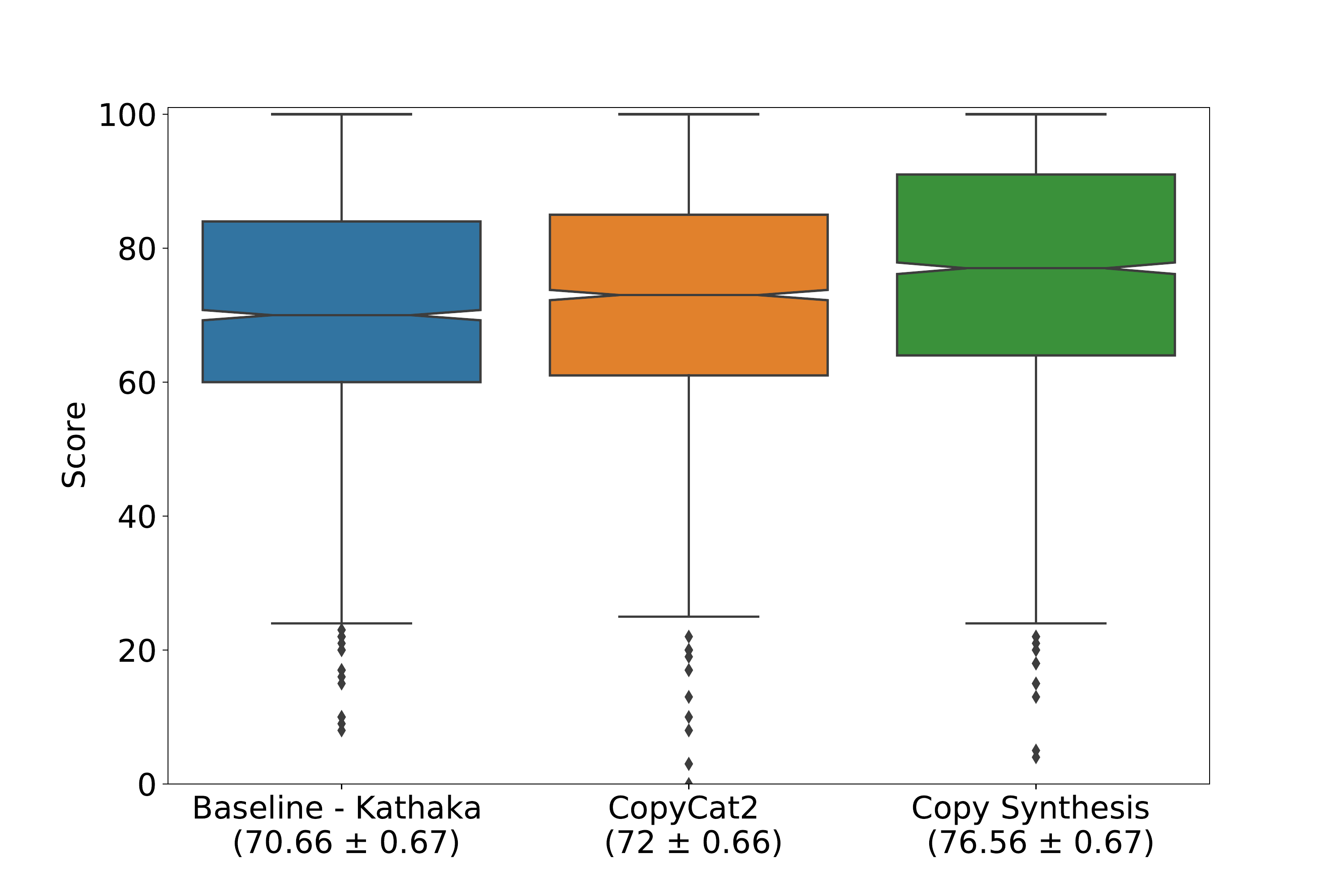}
	        	\vspace{-0.6cm}
				\caption{TTS - Naturalness ($\pmb{22.79\%}$)}
				\label{fig:tts_naturalness}
			\end{subfigure}
			\begin{subfigure}{0.33\linewidth}
				\includegraphics[width=\linewidth]{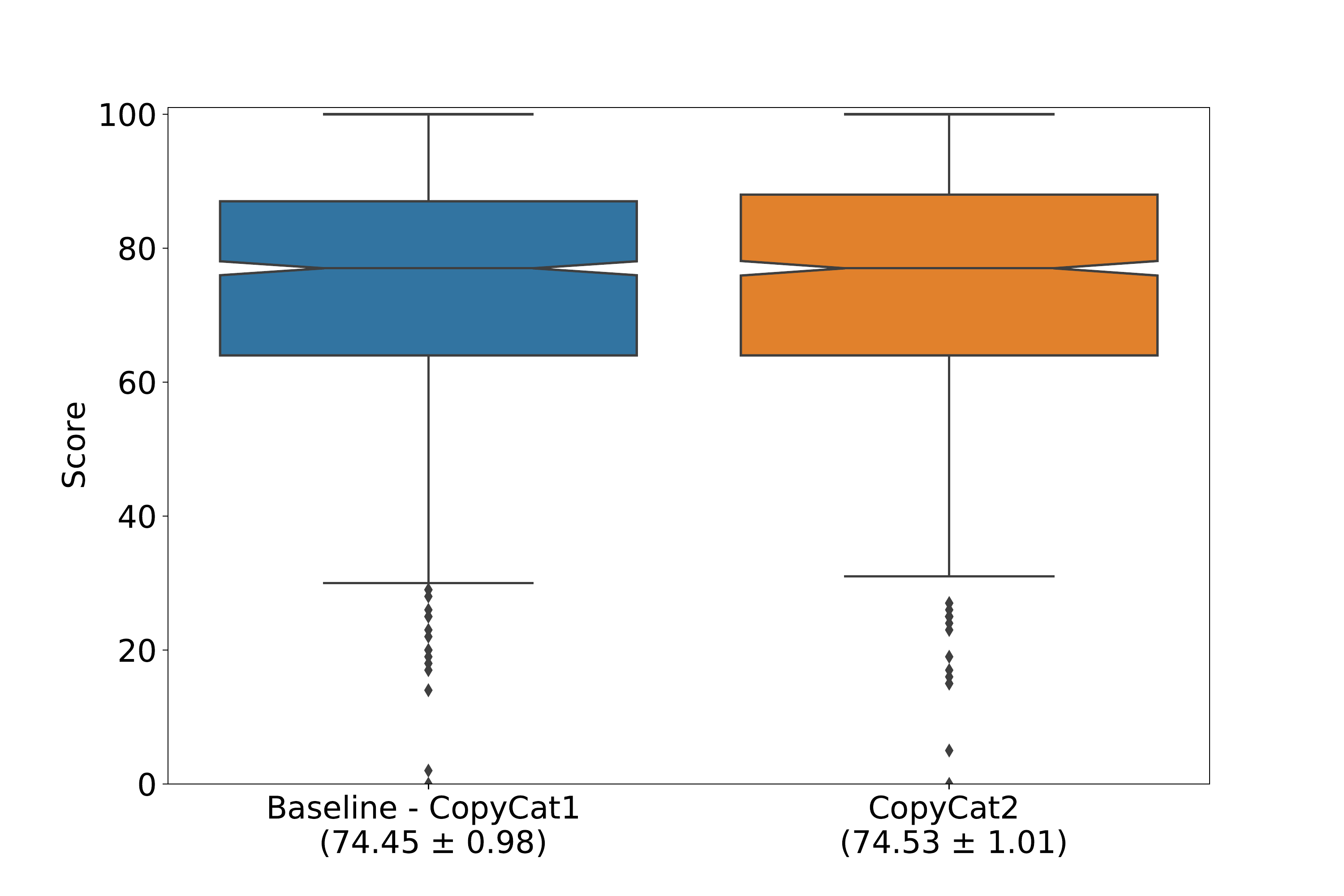}
				\vspace{-0.6cm}
				\caption{Prosody Transfer Quality ($0.29\%$)}
				\label{fig:pt_quality}
			\end{subfigure}
			\begin{subfigure}{0.33\linewidth}
				\includegraphics[width=\linewidth]{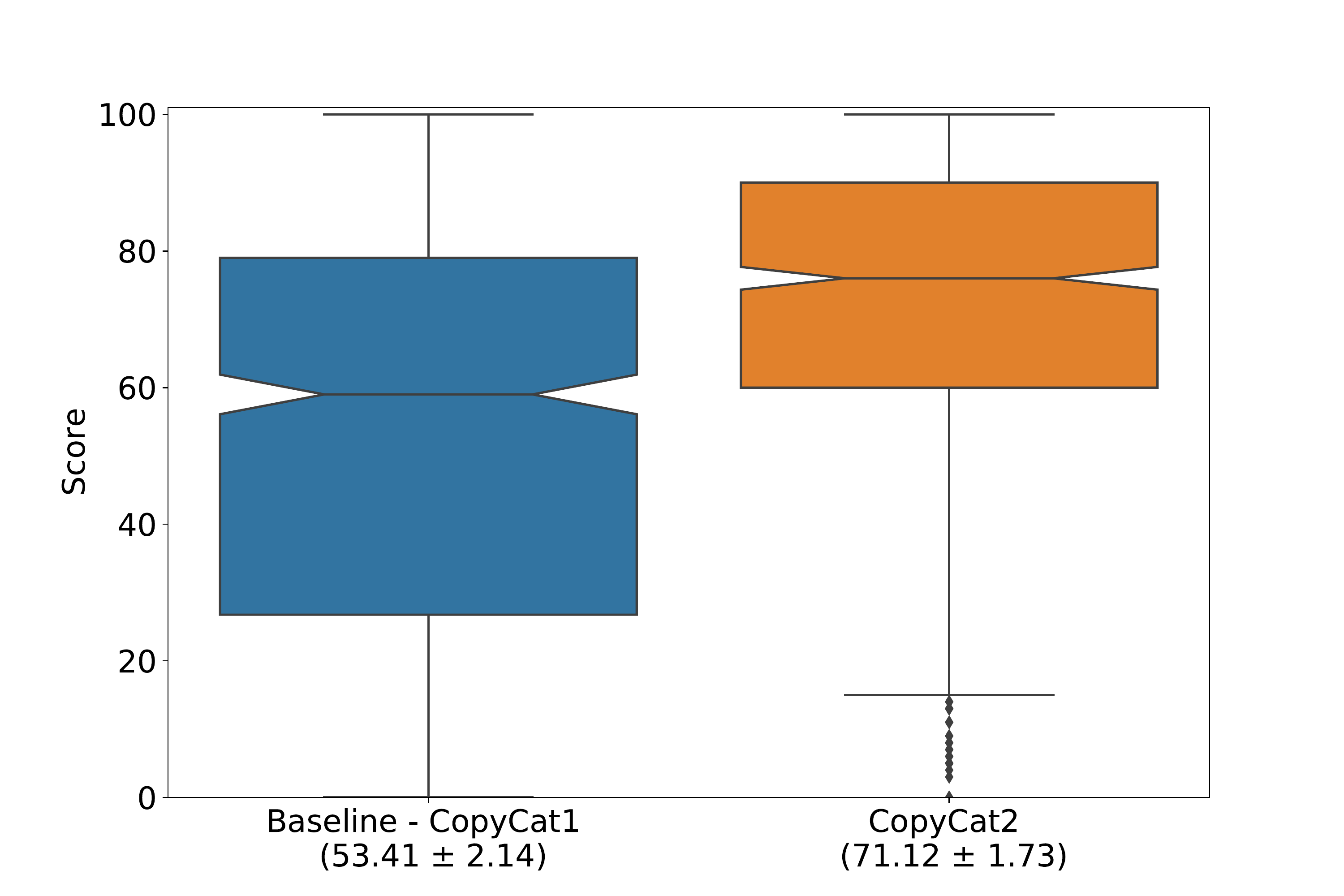}
				\vspace{-0.6cm}
				\caption{Target Speaker Similarity ($\pmb{33.15\%}$)}
				\label{fig:pt_speaker_similarity}
			\end{subfigure}
			\vspace{-.3cm}
			\caption{Here, we show box plots of the scores obtained by each of the systems the MUSHRA evaluations. The average score obtained by each system in the evaluation, along with the $95\%$ confidence interval are shown below each system. The gap reduction obtained by CC2 between the baseline and copy-synthesis in the TTS - Naturalness Evaluation is $22.79\%$. The relative improvements obtained by CC2 when compared to CC1 in the FPT-Transfer Quality and FPT-Speaker Similarity evaluations are $0.29\%$ and $33.15\%$ respectively.}
			\vspace{-.6cm}
		\end{figure*}
		\subsection{Data}
		    \label{ssec:data}
		    We conducted evaluations on an internal multi-speaker US English dataset of varied styles, including, news, facts, greetings, etc. The training dataset consists of a combined total of $50$ hours of speech recorded by $3$ female speakers and $1$ male speaker. Our test dataset consists of an hour of recordings from each of the speakers, covering each of the styles recorded by that speaker. All audio used has a sampling rate of $24$kHz.
		    \vspace{-0.2cm}
        \subsection{Baselines}
            \label{ssec:baselines}
            To evaluate FPT quality and target speaker similarity, we compare our model to CC1, a strong baseline in FPT. We trained CC1 on the same multi-speaker data as CC2. To evaluate the TTS capabilities of our model, we compare it to a strong baseline in representation learning with contextual sampling, Kathaka \cite{Kathaka}. We chose this model as it also follows a similar two-stage training regime, while learning prosody representations from single-speaker data at the sentence-level. It also varies the predicted durations to be more contextually relevant and improves on DurIAN \cite{yu2020durian} in evaluations \cite{Kathaka}. We trained one Kathaka model per speaker in our dataset. 
		\subsection{Evaluations}
			\label{ssec:evaluation}
			\vspace{-0.15cm}
			We conducted $3$ subjective MUSHRA evaluations\cite{itu20031534} to compare CC2 to the TTS and FPT baselines. Each of the evaluations was conducted on all $4$ speakers in the test dataset, and was evaluated by $20$ native US English listeners from a specialized third-party provider. All mel-spectrograms were synthesised using our universal neural vocoder\cite{UniversalWavenet}. In order to negate the effect of any errors introduced by the vocoder, we re-synthesised the recordings using the vocoder and used these copy-synthesised samples in place of human recordings as the upper anchor in the evaluations. We evaluated statistical significance using pairwise two-sided Wilcoxon signed-rank tests.
			\vspace{-0.25cm}
			\subsubsection{FPT - Transfer Quality}
			    \label{sssec:pt_transfer_quality}
			    \vspace{-0.15cm}
			    To evaluate the quality of FPT, we presented each listener with $40$ test cases: $10$ from each of the $4$ target speakers. Each test case consisted of FPT samples from CC1 and CC2, in the target speaker’s voice and the reference prosody sample from one of the other $3$ speakers. The listeners were asked to rate each sample on a scale of $0-100$ based on how closely that sample follows the reference’s prosody: rhythm, emphasis, melody, and loudness of speech. CC2 was on-par with CC1 for all speakers (p-value $ > 0.05$). We found that the average number of mel-spectrogram frames a word spans over was approximately equal to the fixed downsampling rate, $\beta$, of CC1. Therefore, we hypothesise that there was limited change in the amount of fine-grained prosody information encoded by the word-level prosody representations of CC2 and the fixed downsampling rate prosody representations of CC1; thereby, resulting in at-par systems. Additionally, we also saw that altering the durations to match the target speaker's duration distribution while retaining the source's rhythm did not lower the FPT quality.
			    \vspace{-0.25cm}
		    \subsubsection{FPT - Speaker Similarity}
		        \label{sssec:pt_speaker_similarity}
		        \vspace{-0.15cm}
		        To evaluate target speaker similarity, we presented each listener with $40$ test cases: $10$ from each of the $4$ target speakers. Each test case consisted of prosody transferred samples from CC1 and CC2. It also had a reference sample picked randomly from the target speaker's test set to provide the target speaker identity to the listeners. They were asked to rate each sample on how closely it resembled the voice identity of the reference target speaker on a scale of $0-100$. As shown in Fig.\ref{fig:pt_speaker_similarity}, CC2 relatively improves on CC1 by a statistically significant $33.15\%$ (p-value $ < 10^{-4}$). It shows a minimum improvement of $31.17\%$ and a maximum improvement of $35.4\%$ per speaker. We hypothesise this is due to reduced speaker leakage owing to CC2 learning prosody representations at the word-level. In CC1, they are learnt with a fixed downsampling rate, which meant that some of the prosody representations partially covered some linguistic units, resulting in speaker leakage through these representations. In CC2, unlike in CC1, we are certain that the prosody representations capture the prosody of entire words, thereby reducing speaker leakage. We know that speaking rate influences speaker similarity and therefore conjecture that altering the durations to match the target speaker's duration distribution while retaining the source speaker's rhythm could have resulted in increased speaker similarity.
		        \vspace{-0.25cm}
			\subsubsection{TTS - Naturalness}
			    \label{sssec:tts_naturalness}
			    \vspace{-0.2cm}
			    To evaluate the performance of our model in TTS mode, we conducted a $3$-system MUSHRA consisting of audio samples from CC2, single-speaker Kathaka models, and copy-synthesis. We presented each listener with $100$ audio samples picked uniformly at random from the test dataset of each of the $4$ speakers. We asked the listeners to rate each sample based on how natural it sounded, on a scale of $0-100$. As shown in Fig.\ref{fig:tts_naturalness}, on average, CC2 statistically significantly reduces the gap between Kathaka and copy-synthesis by $22.79\%$ in terms of naturalness (p-value $ < 10^{-3}$). It shows a minimum reduction of $19.3\%$ and a maximum reduction of $24.7\%$ per speaker. We hypothesise that training on a multi-speaker dataset at the word-level, helped get denser and fine-grained, acoustic and duration prosody representations. Kathaka learns representations at the sentence-level and from a single speaker dataset, resulting in coarse-grained and sparser representations. This could have also aided the prosody prediction learn a better context-to-prosody mapping in CC2. We speculate that the duration model, which uses predicted duration prosody representations to make the rhythm of speech contextually appropriate, could have further aided in improve CC2's naturalness.
			    \vspace{-0.25cm}
	
	\section{Conclusion}
		\label{sec:conclusion}
		\vspace{-0.1cm}
		In this paper, we presented CopyCat2, and demonstrated $4$ key contributions. We showed that learning prosody representations at the word-level helps improve target speaker similarity in FPT. This was evaluated through a target speaker similarity evaluation, in which CC2 showed a relative improvement of $33.15\%$ against a strong baseline. Through our experiments, we demonstrated that prosody representations learnt for FPT can be used for the downstream task of TTS with contextually appropriate prosody. Using our novel methodology, we demonstrated how we use a single model for both multi-speaker TTS and many-to-many FPT. Finally, we showed that training on multi-speaker data to learn word-level prosody representations and learning to predict them using text, helps improve the naturalness of generated speech. CC2 reduced the gap in naturalness between a strong baseline in TTS with contextually appropriate prosody and copy-sythesised speech by $22.79\%$.

	\newpage
    \bibliographystyle{IEEEtran}
    \bibliography{references}

\begin{thebibliography}{10}
\providecommand{\url}[1]{#1}
\csname url@samestyle\endcsname
\providecommand{\newblock}{\relax}
\providecommand{\bibinfo}[2]{#2}
\providecommand{\BIBentrySTDinterwordspacing}{\spaceskip=0pt\relax}
\providecommand{\BIBentryALTinterwordstretchfactor}{4}
\providecommand{\BIBentryALTinterwordspacing}{\spaceskip=\fontdimen2\font plus
\BIBentryALTinterwordstretchfactor\fontdimen3\font minus
  \fontdimen4\font\relax}
\providecommand{\BIBforeignlanguage}[2]{{%
\expandafter\ifx\csname l@#1\endcsname\relax
\typeout{** WARNING: IEEEtran.bst: No hyphenation pattern has been}%
\typeout{** loaded for the language `#1'. Using the pattern for}%
\typeout{** the default language instead.}%
\else
\language=\csname l@#1\endcsname
\fi
#2}}
\providecommand{\BIBdecl}{\relax}
\BIBdecl

\bibitem{tacotron2}
J.~Shen, R.~Pang, R.~J. Weiss, M.~Schuster, N.~Jaitly, Z.~Yang, Z.~Chen,
  Y.~Zhang, Y.~Wang, R.~Skerrv-Ryan \emph{et~al.}, ``{Natural TTS synthesis by
  conditioning wavenet on mel-spectrogram predictions},'' in \emph{{Proc. of
  IEEE International Conference on Acoustics, Speech and Signal Processing
  (ICASSP)}}, 2018.

\bibitem{UniversalWavenet}
Y.~Jiao, A.~Gabryś, G.~Tinchev, B.~Putrycz, D.~Korzekwa, and V.~Klimkov,
  ``{Universal Neural Vocoding with Parallel Wavenet},'' in \emph{{Proc. of
  IEEE International Conference on Acoustics, Speech and Signal Processing
  (ICASSP)}}, 2021.

\bibitem{li2018close}
N.~Li, S.~Liu, Y.~Liu, S.~Zhao, and M.~Liu, ``{Neural speech synthesis with
  transformer network},'' in \emph{{Proc. of the AAAI Conference on Artificial
  Intelligence}}, 2019.

\bibitem{lorenzo2018robust}
J.~Lorenzo-Trueba, T.~Drugman, J.~Latorre, T.~Merritt, B.~Putrycz,
  R.~Barra-Chicote \emph{et~al.}, ``{Towards Achieving Robust Universal Neural
  Vocoding},'' in \emph{Proc. of Interspeech}, 2019.

\bibitem{zhang2019learning}
Y.-J. Zhang, S.~Pan, L.~He, and Z.-H. Ling, ``{Learning latent representations
  for style control and transfer in end-to-end speech synthesis},'' in
  \emph{{Proc. of IEEE International Conference on Acoustics, Speech and Signal
  Processing (ICASSP)}}, 2019.

\bibitem{skerry2018towards}
R.~Skerry-Ryan, E.~Battenberg, Y.~Xiao \emph{et~al.}, ``{Towards End-to-End
  Prosody Transfer for Expressive Speech Synthesis with Tacotron},'' in
  \emph{{Proc. of the International Conference on Machine Learning}}, 2018.

\bibitem{CAMP}
Z.~Hodari, A.~Moinet, S.~Karlapati, J.~Lorenzo-Trueba, T.~Merritt, A.~Joly,
  A.~Abbas, P.~Karanasou, and T.~Drugman, ``{Camp: A Two-Stage Approach to
  Modelling Prosody in Context},'' in \emph{{Proc. of IEEE International
  Conference on Acoustics, Speech and Signal Processing (ICASSP)}}, 2021.

\bibitem{Kathaka}
S.~Karlapati, A.~Abbas, Z.~Hodari, A.~Moinet, A.~Joly, P.~Karanasou, and
  T.~Drugman, ``{Prosodic Representation Learning and Contextual Sampling for
  Neural Text-to-Speech},'' in \emph{{Proc. of IEEE International Conference on
  Acoustics, Speech and Signal Processing (ICASSP)}}, 2021.

\bibitem{tyagi2020dynamic}
S.~Tyagi, M.~Nicolis, J.~Rohnke, T.~Drugman, and J.~Lorenzo-Trueba, ``{Dynamic
  Prosody Generation for Speech Synthesis Using Linguistics-Driven Acoustic
  Embedding Selection},'' in \emph{Proc. of Interspeech}, 2020.

\bibitem{klapsas2021word}
K.~Klapsas, N.~Ellinas, J.~S. Sung, H.~Park, and S.~Raptis, ``{Word-Level Style
  Control for Expressive, Non-attentive Speech Synthesis},'' in \emph{Proc. of
  the International Conference on Speech and Computer}, 2021.

\bibitem{guo2022unsupervised}
Y.~Guo, C.~Du, and K.~Yu, ``{Unsupervised Word-Level Prosody Tagging for
  Controllable Speech Synthesis},'' \emph{{arXiv preprint arXiv:2202.07200}},
  2022.

\bibitem{ren2021portaspeech}
Y.~Ren, J.~Liu, and Z.~Zhao, ``{PortaSpeech: Portable and High-Quality
  Generative Text-to-Speech},'' in \emph{{Proc. of Advances in Neural
  Information Processing Systems}}, 2021.

\bibitem{hayashi2019pre}
T.~Hayashi, S.~Watanabe \emph{et~al.}, ``{Pre-Trained Text Embeddings for
  Enhanced Text-to-Speech Synthesis},'' in \emph{Proc. Interspeech}, 2019.

\bibitem{karanasou21_interspeech}
P.~Karanasou, S.~Karlapati, A.~Moinet, A.~Joly, A.~Abbas, S.~Slangen,
  J.~Lorenzo-Trueba, and T.~Drugman, ``{A Learned Conditional Prior for the VAE
  Acoustic Space of a TTS System},'' in \emph{Proc. Interspeech}, 2021.

\bibitem{wang2018style}
Y.~Wang, D.~Stanton, Y.~Zhang, R.-S. Ryan, E.~Battenberg, J.~Shor, Y.~Xiao,
  Y.~Jia, F.~Ren, and R.~A. Saurous, ``{Style Tokens: Unsupervised Style
  Modeling, Control and Transfer in End-to-End Speech Synthesis},'' in
  \emph{{Proc. of the International Conference on Machine Learning}}, 2018.

\bibitem{zhang2018learning}
Y.-J. Zhang, S.~Pan, L.~He, and Z.-H. Ling, ``{Learning Latent Representations
  for Style Control and Transfer in End-to-End Speech Synthesis},'' in
  \emph{{Proc. of IEEE International Conference on Acoustics, Speech and Signal
  Processing (ICASSP)}}, 2019.

\bibitem{lee2018robust}
Y.~Lee and T.~Kim, ``{Robust and Fine-Grained Prosody Control of End-to-End
  Speech Synthesis},'' in \emph{{Proc. of IEEE International Conference on
  Acoustics, Speech and Signal Processing (ICASSP)}}, 2019.

\bibitem{klimkov2019finegrained}
V.~Klimkov, S.~Ronanki, J.~Rohnke, and T.~Drugman, ``{Fine-Grained Robust
  Prosody Transfer for Single-Speaker Neural Text-To-Speech},'' in \emph{Proc.
  Interspeech}, 2019.

\bibitem{karlapati2020copycat}
S.~Karlapati, A.~Moinet, A.~Joly, V.~Klimkov, D.~S{\'a}ez-Trigueros, and
  T.~Drugman, ``{CopyCat: Many-to-Many Fine-Grained Prosody Transfer for Neural
  Text-to-Speech},'' in \emph{Proc. Interspeech}, 2020.

\bibitem{kim2020glow}
J.~Kim, S.~Kim, J.~Kong, and S.~Yoon, ``{Glow-TTS: A Generative Flow for
  Text-to-Speech via Monotonic Alignment Search},'' in \emph{Proc. of Advances
  in Neural Information Processing Systems}, 2020.

\bibitem{miao2020flow}
C.~Miao, S.~Liang, M.~Chen, J.~Ma, S.~Wang, and J.~Xiao, ``{Flow-TTS: A
  Non-Autoregressive Network for Text to Speech Based on Flow},'' in
  \emph{{Proc. of IEEE International Conference on Acoustics, Speech and Signal
  Processing (ICASSP)}}, 2020.

\bibitem{vits-kim21f}
J.~Kim, J.~Kong, and J.~Son, ``{Conditional Variational Autoencoder with
  Adversarial Learning for End-to-End Text-to-Speech},'' in \emph{Proc. of the
  International Conference on Machine Learning}, 2021.

\bibitem{jeong21_interspeech}
M.~Jeong, H.~Kim, S.~J. Cheon, B.~J. Choi, and N.~S. Kim, ``{Diff-TTS: A
  Denoising Diffusion Model for Text-to-Speech},'' in \emph{{Proc. of
  Interspeech}}, 2021.

\bibitem{popov2021grad}
V.~Popov, I.~Vovk, V.~Gogoryan, T.~Sadekova, and M.~Kudinov, ``{Grad-TTS: A
  Diffusion Probabilistic Model for Text-to-Speech},'' in \emph{Proc. of the
  International Conference on Machine Learning}, 2021.

\bibitem{sohn2015learning}
K.~Sohn, H.~Lee, and X.~Yan, ``{Learning Structured Output Representation using
  Deep Conditional Generative Models},'' in \emph{{Proc. of Advances in Neural
  Information Processing Systems}}, 2015.

\bibitem{sonderby2016train}
C.~K. S{\o}nderby, T.~Raiko, L.~Maal{\o}e, S.~K. S{\o}nderby, and O.~Winther,
  ``{How to Train Deep Variational Autoencoders and Probabilistic Ladder
  Networks},'' in \emph{{Proc. of the International Conference on Machine
  Learning}}, 2016.

\bibitem{yu2020durian}
C.~Yu, H.~Lu, N.~Hu, M.~Yu, C.~Weng, K.~Xu, P.~Liu, D.~Tuo, S.~Kang, G.~Lei
  \emph{et~al.}, ``{DurIAN: Duration Informed Attention Network for Speech
  Synthesis},'' in \emph{Proc. of Interspeech}, 2020.

\bibitem{Povey_ASRU2011}
D.~Povey, A.~Ghoshal, G.~Boulianne, L.~Burget, O.~Glembek, N.~Goel,
  M.~Hannemann, P.~Motlicek, Y.~Qian, P.~Schwarz, J.~Silovsky, G.~Stemmer, and
  K.~Vesely, ``{The Kaldi Speech Recognition Toolkit},'' in \emph{{Proc. of
  IEEE Workshop on Automatic Speech Recognition and Understanding}}, 2011.

\bibitem{wagner2010experimental}
M.~Wagner and D.~G. Watson, ``{Experimental and Theoretical Advances in
  Prosody: A Review},'' \emph{{Language and Cognitive Processes}}, 2010.

\bibitem{devlin2019bert}
J.~Devlin, M.-W. Chang \emph{et~al.}, ``{BERT: Pre-training of Deep
  Bidirectional Transformers for Language Understanding},'' in \emph{{Proc. of
  the Annual Conference of the North American Chapter of the Association for
  Computational Linguistics (NAACL)}}, 2019.

\bibitem{rogers-etal-2020-primer}
A.~Rogers, O.~Kovaleva, and A.~Rumshisky, ``{A Primer in {BERT}ology: What We
  Know About How {BERT} Works},'' \emph{{Transactions of the Association for
  Computational Linguistics}}, 2020.

\bibitem{itu20031534}
{ITU-R, Recommendation BS}, ``{1534-1,"Method for the Subjective Assessment of
  Intermediate Quality Levels of Coding Systems (MUSHRA)"},''
  \emph{{International Telecommunication Union}}, 2003.

\end{thebibliography}
\end{document}